\documentclass[aps,pra,twocolumn,superscriptaddress,groupedaddress]{revtex4}

\usepackage{amssymb} \usepackage{color,graphicx} \usepackage{amsmath}
\usepackage{amsbsy} \usepackage{amsthm} \usepackage{bbm}
\usepackage{bm,bbm} \usepackage{float} \usepackage{braket}
\usepackage{placeins}
\usepackage[colorlinks=true,citecolor=blue,linkcolor=red,urlcolor=red]{hyperref}
\usepackage[sort&compress]{natbib}
\usepackage{comment}
\usepackage{stackengine}

\newcommand{\bq}{\begin{equation}} \newcommand{\eq}{\end{equation}}
\newcommand{\bqali}{\bq\begin{aligned}}
\newcommand{\eqali}{\end{aligned}\eq}
\newcommand{\bqn}{\begin{equation*}}
\newcommand{\eqn}{\end{equation*}}

\newcommand\Dd[1]{\operatorname{d}^{#1}\!}
\newcommand\D{\operatorname{d}\!}
\renewcommand\k{{\bf k}}

\renewcommand\r{{\bf r}}

\newcommand\kb{k_\text{\tiny B}}

\newcommand\erf{\operatorname{erf}}
\newcommand\com[2]{[#1,#2]}
\newcommand\acom[2]{\{#1,#2\}}
\newcommand\omegac{\omega_\text{\tiny C}}
\newcommand\omegam{\omega_\text{\tiny m}}
\newcommand\gp{g_\phi}
\newcommand\ain{\hat a_\text{\tiny in}}
\newcommand\gammam{\gamma_\text{\tiny m}}
\newcommand\Dp{D_\phi}
\newcommand\omegap{\omega_\phi}

\newcommand\DNSx{\mathcal S_{\text{\tiny V}}(\omega)}
\newcommand\DNSp{\mathcal S_{\text{\tiny R}}(\omega)}
\newcommand\DNSj{\mathcal S_{j}(\omega)}
\newcommand\omegameff{\omega_\text{\tiny m,eff}}
\newcommand\gammameff{\gamma_\text{\tiny m,eff}}
\newcommand\omegapeff{\omega_{\phi,\text{\tiny eff}}}
\newcommand\Dpeff{D_{\phi,\text{\tiny eff}}}
\newcommand\besseli{\operatorname{I}}
\newcommand\etat{\eta_\text{\tiny V}}
\newcommand\etar{\eta_\text{\tiny R}}

\newcommand\xiv{\xi^\text{\tiny V}}
\newcommand\xir{\xi^\text{\tiny R}}

\graphicspath{{../Figures/}}

\begin{document}

\author{Matteo Carlesso}
\email{matteo.carlesso@ts.infn.it}
\affiliation{Department of Physics, University of Trieste, Strada Costiera 11, 34151 Trieste, Italy}
\affiliation{Istituto Nazionale di Fisica Nucleare, Trieste Section, Via Valerio 2, 34127 Trieste, Italy}

\author{Mauro Paternostro}
\affiliation{Centre for Theoretical Atomic, Molecular and Optical Physics, School of Mathematics and Physics,
Queen's University Belfast, Belfast BT7 1NN, United Kingdom}
\affiliation{Laboratoire Kastler Brossel, ENS-PSL Research University, 24 rue Lhomond, F-75005 Paris, France}
\author{Hendrik Ulbricht}
\affiliation{Department of Physics and Astronomy, University of Southampton, SO17 1BJ, UK}
\author{Andrea Vinante}
\affiliation{Department of Physics and Astronomy, University of Southampton, SO17 1BJ, UK}
\author{Angelo Bassi}
\email{bassi@ts.infn.it}
\affiliation{Department of Physics, University of Trieste, Strada Costiera 11, 34151 Trieste, Italy}
\affiliation{Istituto Nazionale di Fisica Nucleare, Trieste Section, Via Valerio 2, 34127 Trieste, Italy}

\title{Non-interferometric test of the Continuous Spontaneous Localization model based on  rotational {optomechanics}}

\date{\today}
\begin{abstract}
{The Continuous Spontaneous Localization (CSL) model is the best known and studied among collapse models, which modify quantum mechanics and identify the fundamental reasons behind the unobservability of quantum superpositions at the macroscopic scale. Albeit several tests were performed during the last decade, up to date the CSL parameter space still exhibits a vast unexplored region. Here, we study and propose an unattempted non-interferometric test aimed to fill this gap. We show that the angular momentum diffusion predicted by CSL heavily constrains the parametric values of the model when applied to a macroscopic object.}
\end{abstract}
 \maketitle

{\section{Introduction}}
Collapse models are  widely accepted as a well-motivated challenge to the quantum superposition principle of quantum mechanics.
 They modify the Schr\"odinger equation by adding non-linear and stochastic terms whose action is negligible on microscopic systems, hence preserving their quantum properties, but gets increasingly stronger on macroscopic ones, inducing a rapid collapse of the wave-function in space \cite{Ghirardi:1986aa,Ghirardi:1990aa,Bassi:2003ab,Bassi:2013aa,Adler:2009aa}. 
The most studied and used collapse model is the Continuous Spontaneous Localization (CSL) model. It is characterised by a coupling rate $\lambda$ between the system and the noise field allegedly responsible for the collapse,  and a typical correlation length $r_C$ for the latter. Ghirardi, Rimini and Weber (GRW) originally set \cite{Ghirardi:1986aa} $\lambda=10^{-16}$\,s$^{-1}$ and $r_C=10^{-7}$\,m. Later, Adler suggested different values  \cite{Adler:2007ab,Adler:2007ac} namely $r_C=10^{-7}$\,m with $\lambda=10^{-8\pm2}$\,s$^{-1}$ and $r_C=10^{-6}$\,m with $\lambda=10^{-6\pm2}$\,s$^{-1}$. This shows that there is no consensus so far on the actual values of the parameters. 

As the CSL model is phenomenological, the values of $\lambda$ and $r_C$ must be eventually determined by experiments. By now there is a large literature on the subject. Such experiments are important because any test of collapse models is a test of the quantum superposition principle. In this respect, experiments can be grouped in two classes: interferometric tests and non-interferometric ones. The first class includes those experiments, which directly create and detect quantum superpositions of the center of mass of massive systems. Examples of this type are molecular interferometry \cite{Eibenberger:2013aa,Hornberger:2004aa,Toros:2017aa,Toros:2018aa} and entanglement experiment with diamonds \cite{Lee:2011aa,Belli:2016aa}. Actually, the strongest bounds on the CSL parameters come from the second class of non-interferometric experiments, which are sensitive to small position displacements and detect CSL-induced diffusion in position \cite{Bahrami:2014aa,Nimmrichter:2014aa,Diosi:2015ac}. Among them, measurements of spontaneous X-ray emission gives the strongest bound on $\lambda$ for $r_C<10^{-6}$\,m \cite{Curceanu:2015aa,Piscicchia:2017aa}, while force noise measurements on nanomechanical cantilevers \cite{Vinante:2016aa,Vinante:2017aa,Carlesso:2018aa} and on gravitational wave detectors give the strongest bound for $r_C>10^{-6}$\,m \cite{Carlesso:2016ac,Helou:2017aa}.

So far research mainly focused on CSL-induced linear diffusion. Very recent technological developments allow to achieve better and better control of rotational motion of non-spherical objects \cite{Kuhn:2017ab,Toros:2018ab,Rashid:2018aa}, thus paving the way to testing rotational CSL-induced diffusion \cite{Collett:2003aa,Bera:2015aa,Schrinski:2017aa}.

By taking the non-interferometric perspective, in this paper we address the potential effects of the CSL mechanisms on an optomechanical system endowed with heterogeneous degrees of freedom. In particular, we consider the roto-vibrational motion of a  {system} coupled to the field of an optical cavity. By addressing {its}  ensuing  dynamics, we show that the rotational degree of freedom offers enhanced possibilities for exploring a wide-spread region of the parameters space of the CSL model, thus contributing significantly to the ongoing quest for the validity of collapse theories. We provide a thorough assessment of the experimental requirements for the envisaged test to be realized and highlight the closeness of our proposal to state-of-the-art experiments. \\

{\section{Theory}}

{
In order to fix the ideas, here we focus on an optomechanical setup whose vibrational and rotational degrees of freedom are monitored. Although the range of masses spanned by typical optomechanical experiments is very large~\cite{Aspelmeyer:2014aa} (from the zg scale of atomic gases \cite{Murch:2008aa} to the 40\,kg of the LIGO mirrors \cite{Abbott:2016ab}), the measurement technique is conceptually the same for all cases and it is commonly performed by means of the coupling to an optical mode, which is then read out to infer the noise properties of the mechanical system. Specifically, the density noise spectrum of suitable observables of the optical mode is typically used as the workhorse to gather insight into the 
motion of the mechanical system~\cite{Paternostro:2006aa}. As illustrated later on in this Section, this will embody our detection scheme for the roto-vibrational motion that we focus on.}\\

\noindent
\textit{Optomechanical setup -- }
\begin{figure}[t!]
\includegraphics[width=0.6\linewidth]{{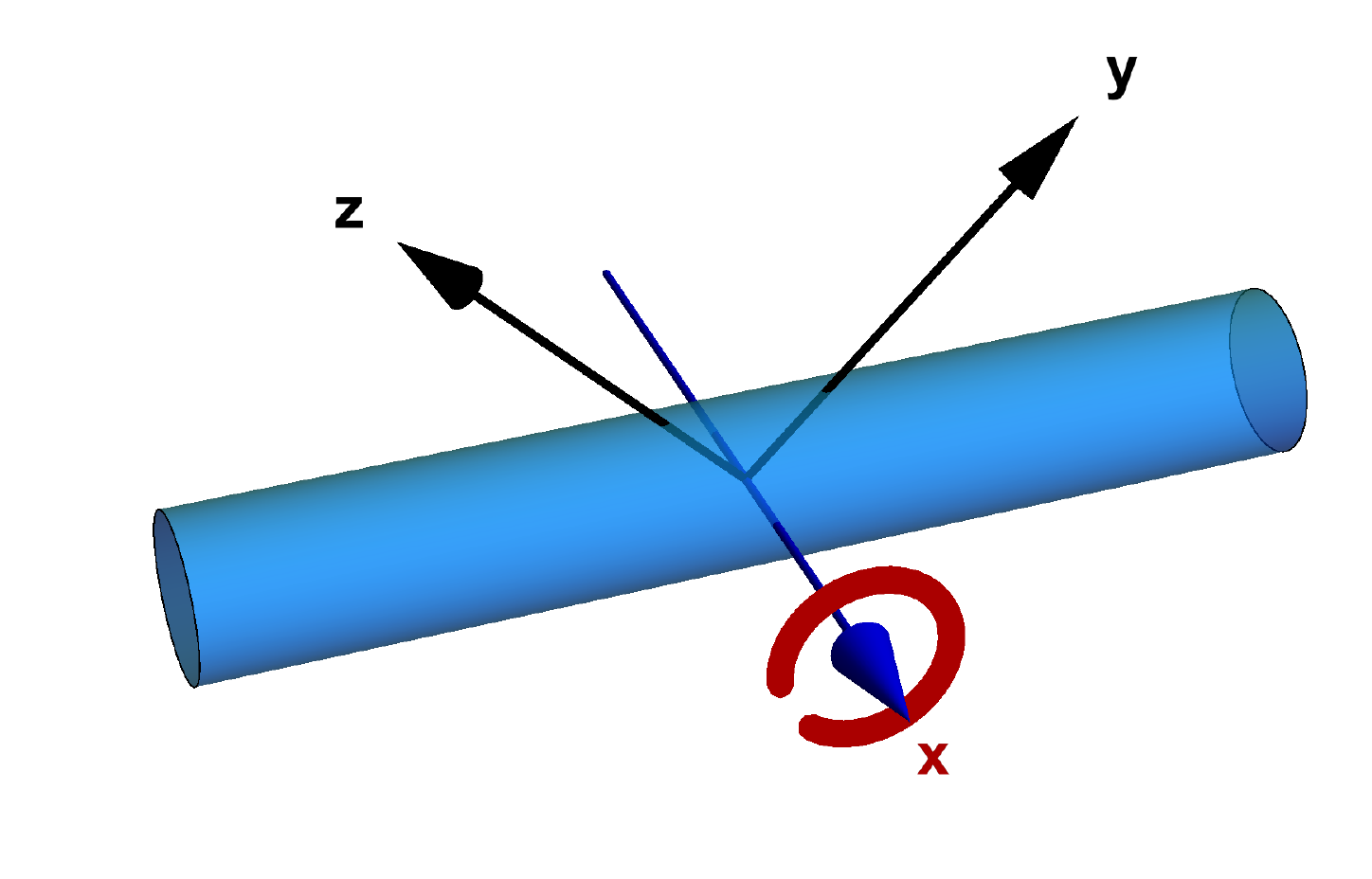}}
\caption{(Color online) Graphical representation of the cylinder with respect to the chosen Cartesian axes. The monitored motions are the vibration along the $x$ axis and the rotation around it (represented in blue and red respectively).}
\label{fig:rod}
\end{figure}
{Let us consider} a cylinder harmonically trapped, both in position and in angle,{ whose} monitored motions are the center of mass vibrations along the $x$ axis and the rotations around it. We consider the system symmetry axis oriented orthogonally to the direction of light propagation (cf.~Fig.~\ref{fig:rod}).
The Hamiltonian, describing the vibrational motion of the cylinder harmonically trapped at frequency $\omegam$ by interacting with a cavity field, is given by \cite{Mancini:1994aa,Walls:2007aa}
\bq\label{eq.H.vibr}
\hat H_\text{V}=\hbar\omegac\hat a^\dag\hat a+\frac{\hat p^2}{2m}+\frac12m\omegam^2\hat x^2-\hbar\chi\hat a^\dag\hat a\hat x.
\eq
In Eq.~\eqref{eq.H.vibr} the first term describes the free evolution of the cavity mode at frequency $\omegac$, with $\hat a^\dag$ and $\hat a$ denoting the photons' creation and annihilation operators; the next two terms describe the oscillatory motion of the mass $m$ in the cavity, where $\hat x$ and $\hat p$ are, respectively, its position and momentum operators. The last term describes the interaction between the cavity field and the vibrational motion of the system, with coupling constant $\chi$.
If we include also rotations of the cylinder along the direction of propagation of the radiation field ($x$ axis) we need to consider the additional term \cite{Bhattacharya:2007aa,Bhattacharya:2008aa}
\bq\label{eq.H.rot}
\hat H_\text{R}=\frac{\hat L_x^2}{2I}+\frac12I\omegap^2\hat\phi^2-\hbar\gp\hat a^\dag\hat a\hat \phi,
\eq
which is the rotational Hamiltonian, characterized by a moment of inertia $I$ and torsional trapping frequency $\omegap$; the third term, proportional to the coupling constant $\gp$, accounts for the laser interaction with the rotational degrees of freedom. In Eq.~\eqref{eq.H.rot}, $\hat \phi$ is the angular operator describing rotations along the $x$ axis, such that $\com{\hat \phi}{\hat L_x}=i\hbar$, with $\hat L_x$ the angular momentum operator along the same direction.\\

\noindent
\textit{CSL Model -- }
The master equation describing the evolution of the density matrix of a system affected by a CSL-like mechanism \cite{Bassi:2003ab,Bassi:2013aa} is of the Lindblad form $\partial_t\hat \rho=-\tfrac i\hbar\com{\hat H}{\hat \rho}+\mathcal L[\hat \rho]$, where $\hat H$ is the free Hamiltonian of the system and
\bq\label{csl.master}
\mathcal L[\hat \rho]=-\frac{\lambda}{2r_C^3\pi^{3/2}m_0^2}\int\Dd{3} \r\,\com{\hat M(\r)}{\com{\hat M(\r)}{\hat \rho}},
\eq
with $\hat M(\r)=\sum_n m_n\exp({-{(\r-\hat\r_n)^2}/{2r_C^2}})$ and $\hat \r_n$ the position operator of the $n$-th nucleon (of mass $m_n$) of the system.
 Under the approximation of small fluctuations of the center-of-mass of a rigid object and small rotations of the system under the action of the CSL noise, two conditions that are fulfilled in typical opto-mechanical setups, $\mathcal L[\hat \rho]$ can be Taylor expanded around the equilibrium position. The center of mass motion and the system's rotations can be decoupled from the internal dynamics and Eq.~\eqref{csl.master} reduces to \cite{Schrinski:2017aa}
\bq\label{master.eq}
\mathcal L[\hat \rho]\simeq-\frac\etat2\com{\hat x}{\com{\hat x}{\hat \rho}}-\frac{\etar}{2}\com{\hat \phi}{\com{\hat \phi}{\hat \rho}},
\eq
which represents an extension of the master equation describing only the pure center of mass vibrations (the first of the two terms) to the roto-vibrational case; its general form for an arbitrary geometry of the system can be found in \cite{Schrinski:2017aa}. The explicit forms of the vibrational ($\etat$) and rotations ($\etar$) diffusion constants are reported in Appendix \ref{AppB}.

{Clearly, Eq.~\eqref{master.eq} predicts a diffusion of the linear and angular momentum and optomechanical setups are ideal sensors to measure such effects. There is a large variety employed in these experiments and external influences can be monitored very accurately. }

The corresponding equations of motion can be obtained by merging the Hamiltonian optomechanical dynamics in Eqs.~\eqref{eq.H.vibr} and~\eqref{eq.H.rot} and the CSL-induced diffusions described by Eq.~\eqref{master.eq}, to which we add the dampings and thermal noises \cite{Gardiner:2004aa}. Explicitly, we get the equations \cite{Bhattacharya:2007aa}: $\D{\hat x}/\D t={\hat p/m}$, $\D{\hat \phi}/\D t={\hat L_x/I}$ and
\bqali\label{langevin.eq}
\frac{\D{\hat a}}{\D t}&=-i(\Delta_0-\gp\hat\phi-i \kappa) \hat a+i\chi\hat a\hat x+\sqrt{2\kappa}\ain,\\
\frac{\D{\hat p}}{\D t}&=-m\omegam^2\hat x+\hbar\chi\hat a^\dag\hat a-\gammam \hat p+\hat \xiv-\hbar\sqrt{\etat} w_\text{\tiny V},\\
\frac{\D{\hat L_x}}{\D t}&=-I\omegap^2\hat \phi+\hbar \gp\hat a^\dag\hat a-\frac{\Dp}{I}\hat L_x+\hat\xir-\hbar\sqrt{\etar} w_\text{\tiny R},
\eqali
where $\Delta_0=\omegac-\omega_0$ is the detuning of the laser frequency $\omega_0$ from the cavity resonance; $\kappa$, $\gammam$ and $\Dp$ are the damping rates for the cavity, for the vibrations and rotations of the system respectively; $\ain$ is a noise operator describing the incident laser field, defined by the input power $P_\text{\tiny in}=\hbar\omegac|\alpha|^2$, with $\alpha=\braket{\ain}$, and delta-correlated fluctuations $\braket{\delta\ain(t)\delta\ain^\dag(s)}=\delta(t-s)$, where $\ain=\alpha+\delta\ain$. The noise operators $\hat \xiv$ and $\hat \xir$ describe the thermal action of the surrounding environment (supposed to be in equilibrium at temperature $T$), which is assumed to act independently on vibrations and rotations. They are assumed to be Gaussian with zero mean and correlation function \cite{Bhattacharya:2007aa}
\bq
\label{twopoint}
\frac{\braket{\hat\xi^{j}_t\hat \xi^j_s}}{\hbar\epsilon_j}
=\int\frac{\D\omega}{2\pi}e^{-i\omega(t-s)}\omega[1+\coth(\beta\omega)]~~(j=\text{R,V}),
\eq
with $\beta=\hbar/2\kb T$, $\epsilon_\text{\tiny R}= D_\phi$, and $\epsilon_\text{\tiny V}=m\gamma_m$.
As already discussed, the CSL noise acts as a source of stochastic noise, whose influence on the dynamics of the system is encompassed by the addition, in Eqs.~\eqref{langevin.eq}, of the force terms $-\hbar\sqrt{\eta_j} w_j~(j=\text{R,V})$ with $\braket{w_j}=0$ and $\braket{w_i(t)w_j(s)}=\delta_{ij}\delta(t-s)$~\cite{Bahrami:2014aa,Nimmrichter:2014aa}. 

From Eqs.~\eqref{langevin.eq} and \eqref{twopoint} we can derive the density noise spectrum (DNS) associated to $\delta \tilde x(\omega)$ and $\delta \tilde\phi(\omega)$, which are the fluctuations of the position and angle operators in Fourier space respectively, $\DNSj=\tfrac{1}{4\pi}\int\D\Omega\braket{\acom{\delta \tilde O_j(\omega)}{\delta \tilde O_j(\Omega)}}$~$(j=\text{V,R})$. 
\begin{table}[t!]
\begin{tabular}{c|ccccc}
\hline\hline
\ DNS Parameter\ \ &\ ${\cal G}_j$& $\omega_{j,\text{eff}}$&$\Gamma_{j,\text{eff}}$&$\lambda_j$ \\ 
 \hline
{ Vibration}& $\chi$ &$\omegameff$ &$\gammameff$  & $m$\\ 
 { Rotation}&  $g_\phi$& $\omegapeff$& $\Dpeff/I$ & $I$\\
\hline\hline
\end{tabular}
\caption{\label{tavola}Explicit form of the parameters entering the DNS of the fluctuations of the rotational and vibrational degrees of freedom of the system [cf. Eq.~\eqref{dns.eq}].}
\end{table}

Through a 
lengthy but straightforward calculation, the explicit form of both $\DNSx$ and $\DNSp$ can be calculated and put under the Lorentzian form
\begin{widetext}
\bq\label{dns.eq}
\DNSj=\frac{2\hbar^2|\alpha|^2\kappa {\cal G}^2_j+\left[\kappa^2+(\Delta-\omega)^2\right]\left[	\hbar \omega\epsilon_j\coth(\beta\omega) +\hbar^2\eta_{j}\right]}{\lambda_j^2\left[\kappa^2+(\Delta-\omega)^2\right][(\omega_{j,\text{eff}}^2-\omega^2)^2+\Gamma_{j,\text{eff}}^2\omega^2]}.
\eq
\end{widetext}
The parameters specific of the considered degree of freedom that appear in Eq.~\eqref{dns.eq} (${\cal G}_j,\omega_{j,\text{eff}},\Gamma_{j,\text{eff}}$ and $\lambda_j$) are given in Table~\ref{tavola}.
\begin{figure*}[ht!]
\def\bigc{\includegraphics[width=0.33\linewidth]{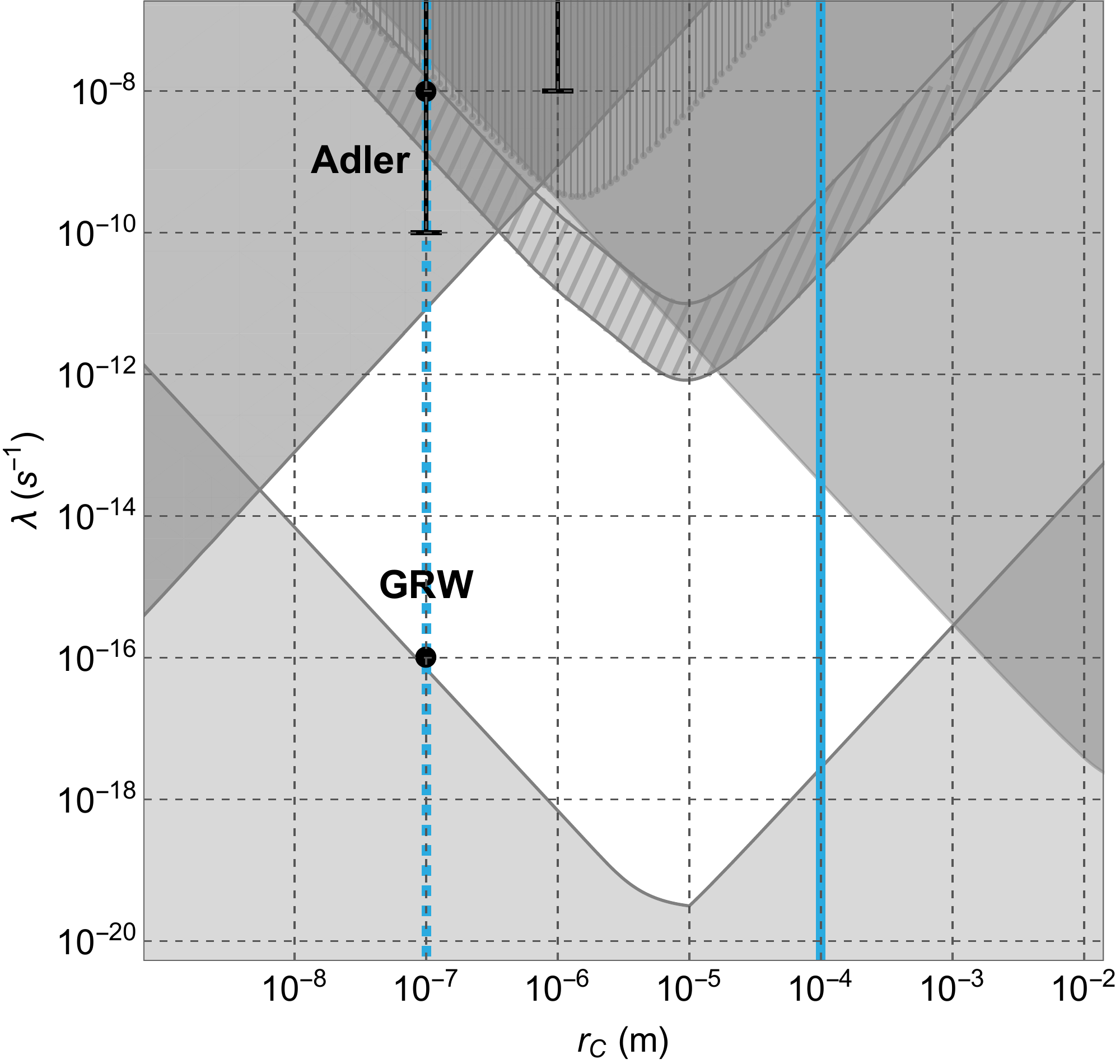}}
\def\bigd{\includegraphics[width=0.33\linewidth]{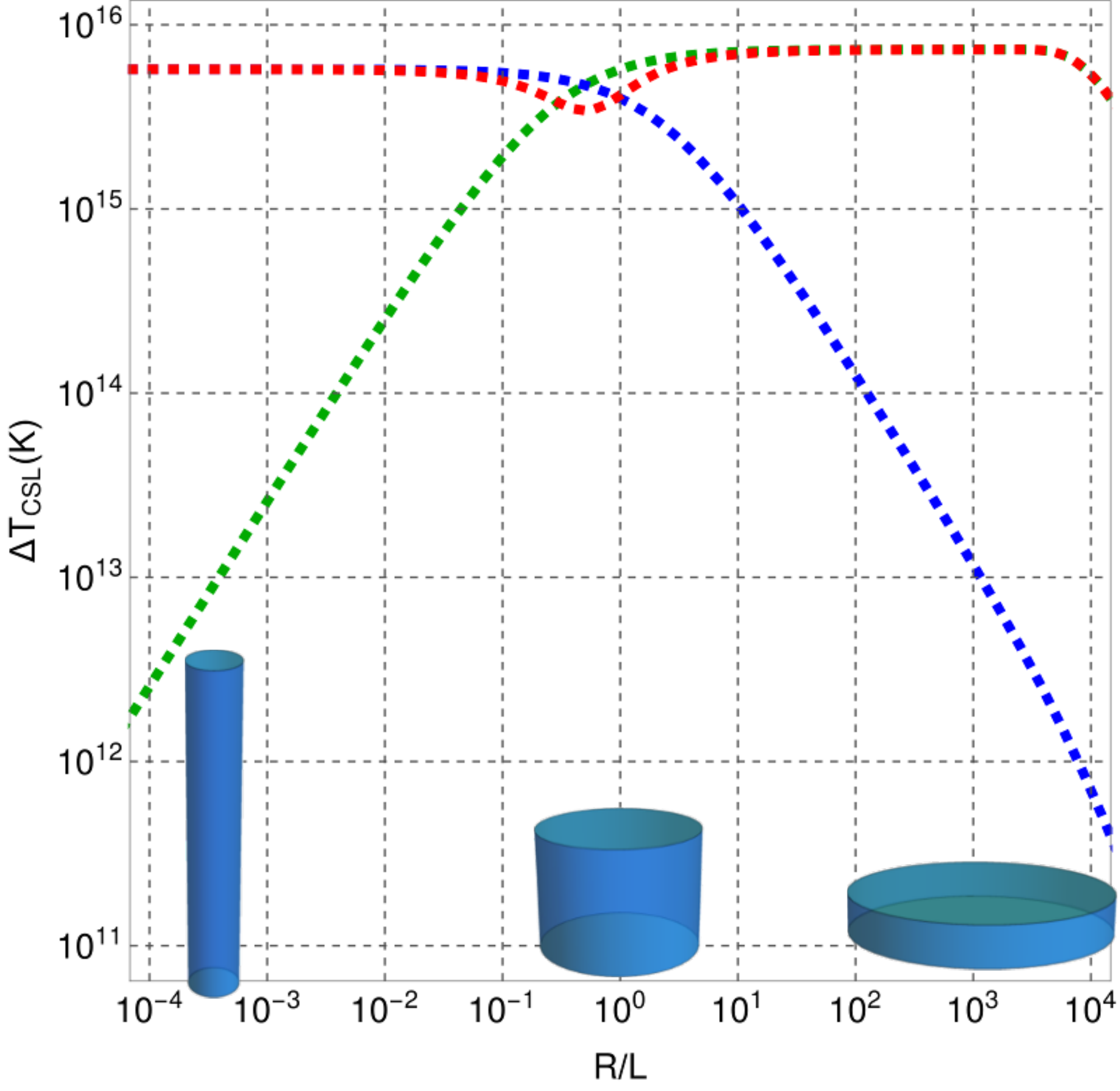}}
\def\littlea{\includegraphics[height=0.35cm]{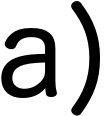}}
\def\littleb{\includegraphics[height=0.35cm]{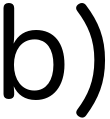}}
\def\littlec{\includegraphics[height=0.35cm]{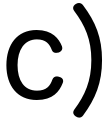}}
\def\lisapic{\includegraphics[width=0.33\linewidth]{{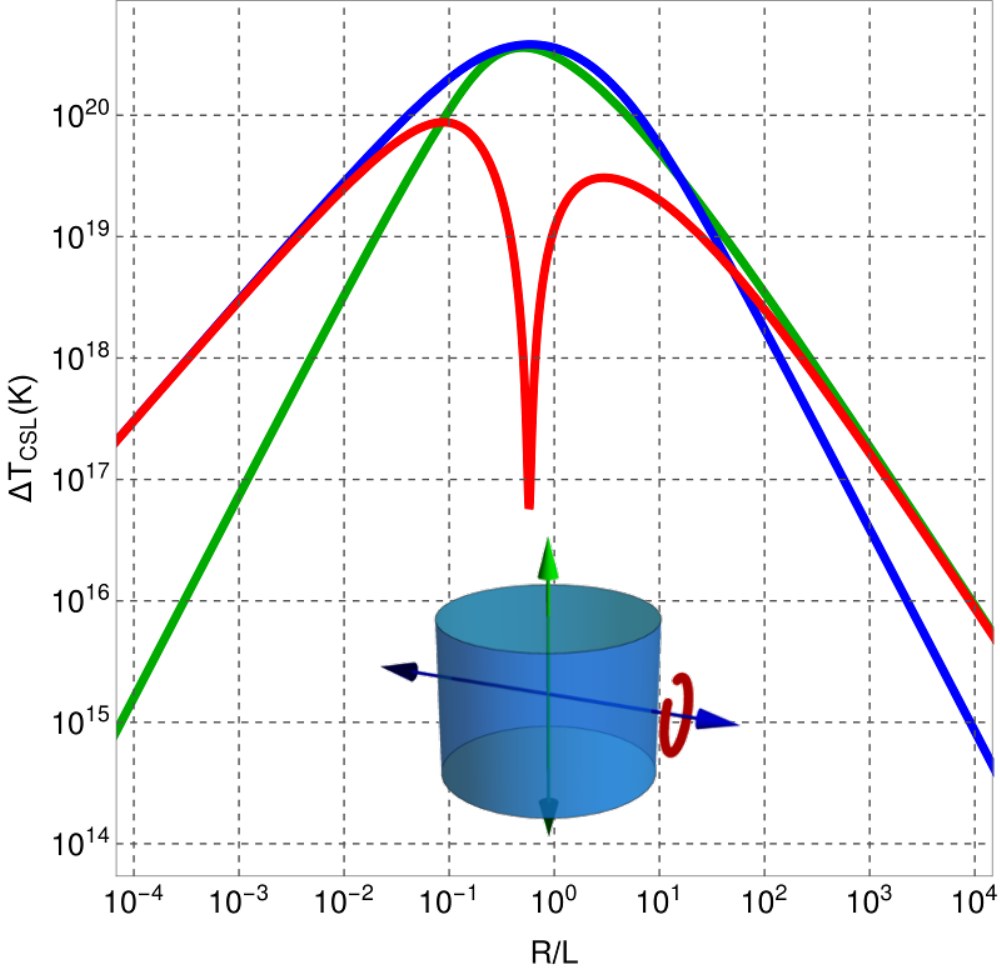}}}
\def\stackalignment{l}
\bottominset{\littlea}{\bigc}{0pt}{10pt}\bottominset{\littleb}{\bigd}{0pt}{10pt}\bottominset{\littlec}{\lisapic}{0pt}{10pt}
\caption{(Color online) Panel \textbf{(a)}: In grey we show the strongest bounds presently reported in literature~\cite{Curceanu:2015aa,Vinante:2016aa,Carlesso:2016ac,Belli:2016aa,Bilardello:2016aa,Laloe:2014aa,Toros:2018aa,Toros:2017aa}. The values suggested by GRW \cite{Ghirardi:1986aa,Ghirardi:1995aa} and Adler \cite{Adler:2007aa}, and the associated ranges, are indicated in black. The cyan lines show the values of $r_C$ we consider in our analysis, namely $r_C=10^{-7}$\,m (dotted line) and $10^{-4}$\,m (continuous line). 
Panels \textbf{(b)} and \textbf{(c)}: CSL temperature contribution $\Delta T_\text{\tiny CSL}$ against $R/L$, for $\lambda=1$\,s$^{-1}$, $r_C=10^{-7}$\,m [panel \textbf{(b)}] and $r_C=10^{-4}$\,m [panel \textbf{(c)}]. The blue and green lines (either dotted or solid) denote the behavior of $\Delta T_\text{\tiny CSL}^\text{\tiny V}$ along the $x$ axis and the symmetry axis, respectively. The red lines (dotted and solid) show $\Delta T_\text{\tiny CSL}^\text{\tiny R}$. The dip in the red curve occurs when the dimensions of the cylinder are similar, which makes it less sensitive to rotations.}
 \label{fig:bestgeom}
\end{figure*} 

We have introduced the effective frequencies $\omegameff$ and $\omegapeff$ and  damping constants $\gammameff$ and $\Dpeff$~\cite{Bhattacharya:2007aa,Genes:2008aa}, whose explicit expressions are presented in Appendix~\ref{AppA}, and $\Delta=\Delta_0-g_\phi\braket{\hat \phi}-\chi\braket{\hat x}$. The CSL contributions are encompassed by the diffusion constant $\eta_j$, which enters $\DNSj$ as an additional heating term akin to the environment-induced one $\hbar \omega\epsilon_j\coth(\beta\omega)$. In the high temperature limit ($\beta\to0$), which is in general valid for typical low-frequency optomechanical experiments, the latter takes the form $\hbar\epsilon_j/\beta$. Therefore, in such a limit, we have
\begin{equation}
\hbar \omega\epsilon_j\coth(\beta\omega) +\hbar^2\eta_{j}\to{\hbar\epsilon_j}\left(\frac{1}{\beta}+\frac{\hbar\eta_j}{\epsilon_j}\right)\equiv\frac{\hbar\epsilon_j}{\beta_{j,\text{eff}}},
\end{equation}
where we have defined the $j$-dependent effective inverse temperature $\beta_{j,\text{eff}}$, thus showing that the different degrees of freedom of the system thermalise to different, in principle distinguishable, CSL-determined temperatures.  This means that CSL gives the extra temperatures
\bq\label{tempcsl}
\Delta T^\text{\tiny V}_\text{\tiny CSL}=\frac{\hbar^2\etat}{2\kb m \gammam}\quad \text{and}\quad \Delta T^\text{\tiny R}_\text{\tiny CSL}=\frac{\hbar^2\etar}{2\kb \Dp}.
\eq
The first was extensively studied both theoretically \cite{Collett:2003aa,Adler:2005aa,Bahrami:2014aa, Nimmrichter:2014aa, Diosi:2015ab, Goldwater:2016aa} and experimentally \cite{Belli:2016aa,Vinante:2016aa, Carlesso:2016ac}. However, for the rotational degree of freedom, the existence of $\Delta T^\text{\tiny R}_\text{\tiny CSL}$ opens up new possibilities for testing the CSL model, as discussed below.\\

  {\section{Lab-based Experiments}}

Upon subtracting the optomechanical contribution to the temperature embodied by the first term in Eq.~\eqref{dns.eq}~\footnote{The contribution arising from the driving field can be accurately calibrated experimentally due to of the relatively 
large intensity of the field (which also makes any uncertainty negligible with respect to the nominal signal). 
Such well characterized contribution can then be subtracted from the density noise spectrum 
to let the features linked to the mechanical motion emerge. {{}{We also note that the optomechanical contribution can be strongly suppressed by implementing a stroboscopic measurement strategy.}}}, the experimental measurement of the temperature of the system is given by $T_\text{\tiny m}\pm \delta T$, where $\delta T$ is the experimental  measurement accuracy. Unless one sees an excess temperature of unknown origin \cite{Vinante:2017aa}, the outcome of the experiment will be $\Delta T_\text{\tiny CSL}\leq\delta T$, thus setting a bound on the collapse parameters  {once Eq.~\eqref{tempcsl} is considered}. We first compare the magnitude of the two temperatures $\Delta T_\text{\tiny CSL}^\text{\tiny V}$ and $\Delta T_\text{\tiny CSL}^\text{\tiny R}$ for different geometries of the system. Without loss of generality (as $\Delta T_\text{\tiny CSL}^\text{\tiny V,R}\propto \eta_j\propto\lambda$) we set $\lambda=1$\,s$^{-1}$. 
For definiteness we take a silica cylinder with $m=10\,\mu$g and vary the ratio between the radius $R$ and the length $L$. For the residual gas, we consider He-4, at the temperature of $T=1$\,K and {{}{pressure $P =5\times 10^{-13}$\,mbar, which can be reached with existing technology \cite{Gabrielse:1990aa}. }}Fig.~\ref{fig:bestgeom} shows the behaviour of $\Delta T_\text{\tiny CSL}^\text{\tiny V,R}$ for $r_C=10^{-7}$\,m (dotted lines) and $r_C=10^{-4}$\,m (continuous lines) \footnote{These values of $r_C$ are chosen due to their closeness to the boundaries of the unexplored CSL parameter region.}. In the latter case the strongest contribution comes from vibrations along the $x$ axis (blue lines) at $2R\sim L$, while in the former case it comes from rotations (red lines) for $R\gg L$  {and $R\ll L$}, thus showing that CSL tests based on rotational motion can be as good or better than those based on vibrational motion.  {For the following analysis we focus on the $R\gg L$ case, which gives the strongest contribution for the originally chosen value of the correlation length $r_C=10^{-7}\,$m.}  As a comparison, we also report $\Delta T_\text{\tiny CSL}$ given by vibrations along the symmetry axis (green lines).

\begin{figure*}[t!]
\def\bigc{\includegraphics[width=0.33\linewidth]{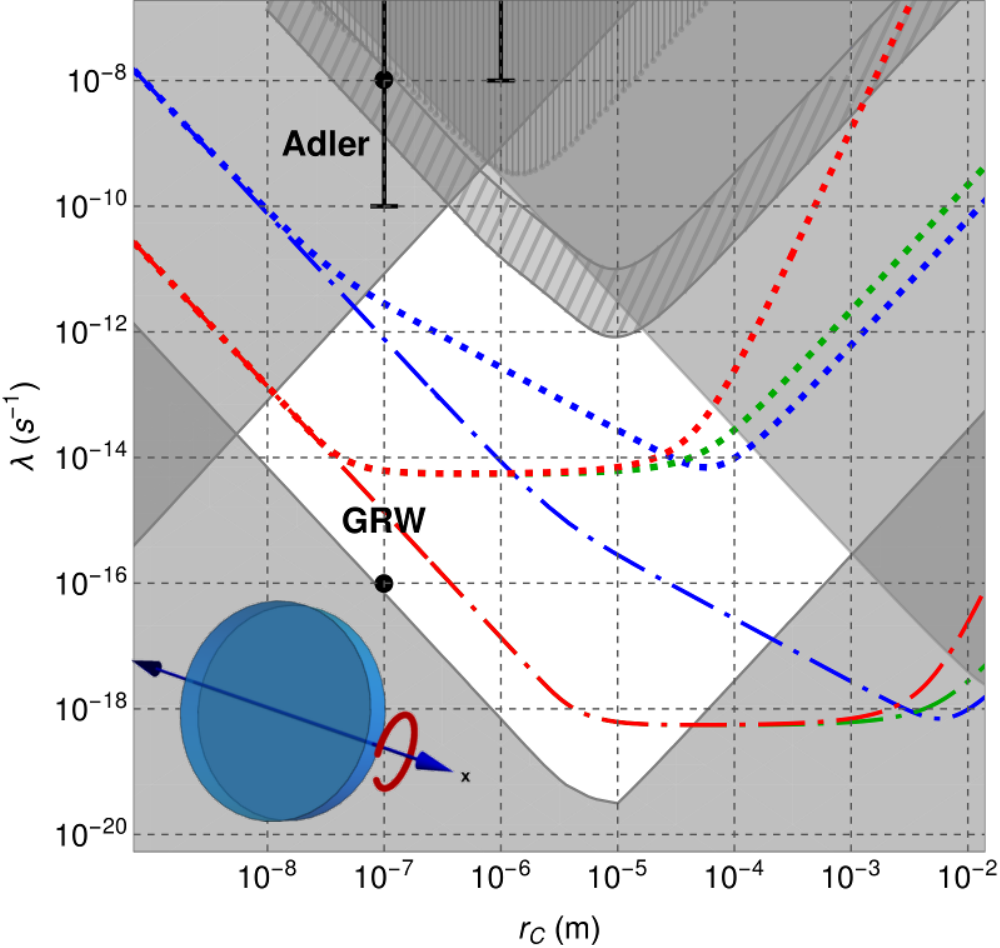}}
\def\littlea{\includegraphics[height=0.35cm]{aaa.png}}
\def\littleb{\includegraphics[height=0.35cm]{bbb.png}}
\def\lisapic{\includegraphics[width=0.33\linewidth]{{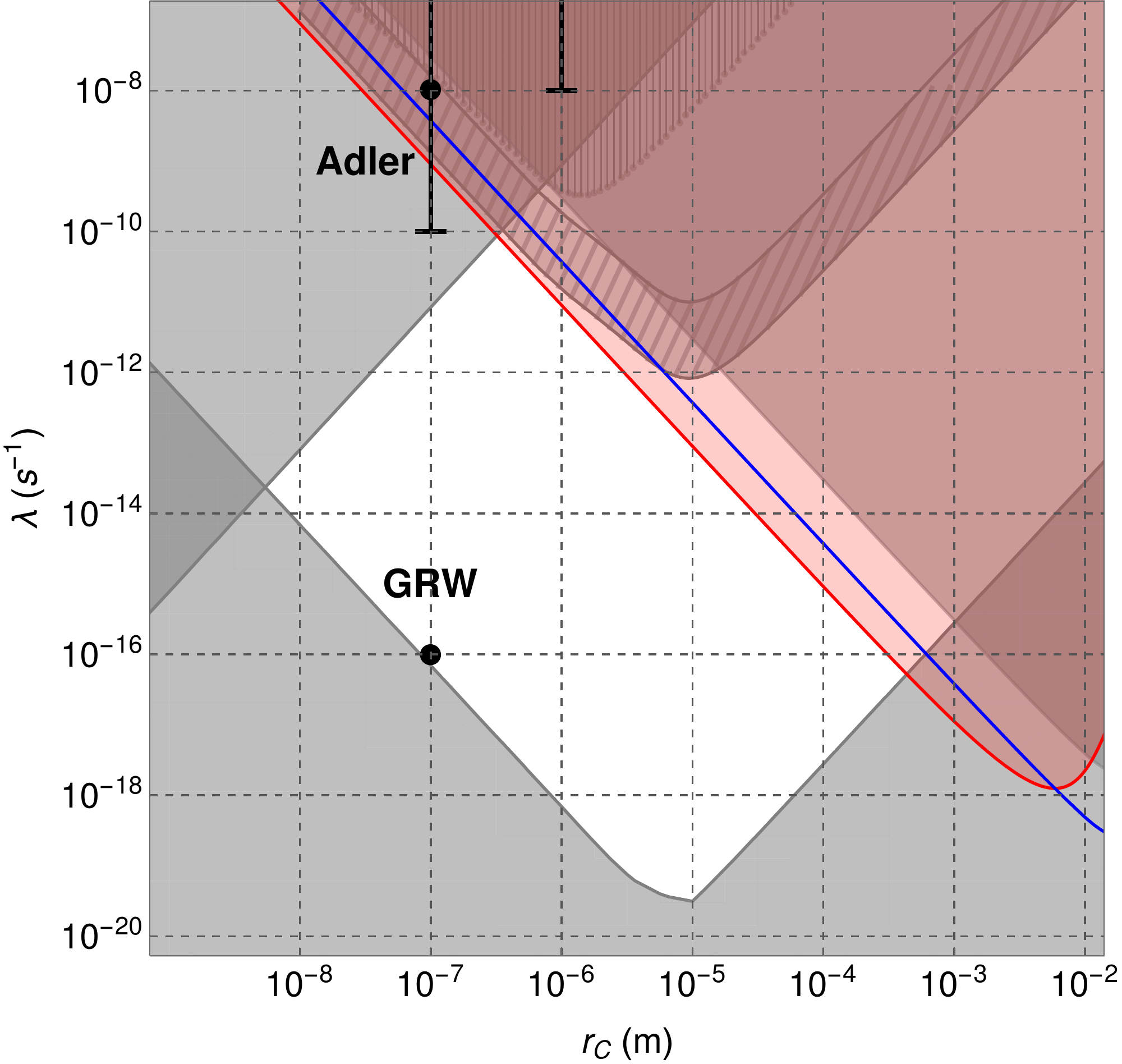}}}
\def\stackalignment{l}
\bottominset{\littlea}{\bigc}{0pt}{10pt}\bottominset{\littleb}{\lisapic}{0pt}{10pt}
\caption{(Color online) Panel \textbf{(a)}: Hypothetical upper bounds obtained assuming $\Delta T_\text{\tiny CSL}\leq\delta T=0.1$\,K.
 The chosen system is a silica cylinder of radius $R$ and length $L$, cooled at the temperature of $T=1$\,K and at a pressure of $P=5\times10^{-13}$\,mbar. Blue, green and red lines: Upper bounds for the vibrational motion along the $x$ axis, the symmetry axis and rotational motion around it respectively. The dotted (dotted-dashed) lines correspond to $R=0.1$mm and $L=0.1\mu$m ($R=1$cm and $L=10\mu$m).
 Panel {{}{\textbf{(b)}: Hypothetical upper bound and exclusion region (red line and region) from the analysis of possible rotational noise from LISA Pathfinder. Blue line: upper bound from LISA Pathfinder improved measurements \cite{Armano:2018aa}, derived as in \cite{Carlesso:2016ac}.}}}
\label{fig:result}
\end{figure*} 

In Fig.~\ref{fig:result}a) we compare the hypothetical upper bounds obtained from the vibrational and rotational motion, taken individually. This is done by setting the accuracy in temperature to $\delta T=0.1$\,K and varying the dimensions of the cylinder.
As a case-study, we consider a thought experiment aimed at testing CSL in the region $r_C\sim10^{-7}$\,m, and exploit rotations of a \emph{coin} shaped system to maximize the CSL effect [cf.~Fig.~\ref{fig:bestgeom}b)].
As shown, the hypothetical upper bound given by the rotational motion is comparable with the vibrational one. \\

\noindent
\textit{Experimental feasibility --} Having assessed formally how, in extreme vacuum conditions, the rotational motion of a levitated cylinder of fairly macroscopic dimensions can set very strong bounds on the CSL model (almost testing the GRW hypothesis), we now address the experimental feasibility of our proposal, showing that the proposed experiment is entirely within the grasp of current technology.  

First, the cylinder has to be trapped magnetically or electrically to allow for its rotational motion around the $x$-axis, as in Fig.~\ref{fig:rod}. Needless to say, we must avoid competing heating effects such those due to gas collisions and exchange of thermal photons between the environment and the trapped cylinder. Moreover, one must ensure the ability to control and detect precisely enough the rotational motion of the trapped cylinder. The first condition can be granted by performing the experiments at low temperatures and pressures.  {{Standard dilution cryostats reach temperatures $<10$\,mK \cite{Vinante:2016aa, Vinante:2017aa} and 
the reachable pressure is as low as $10^{-17}$mbar (as done in the cryogenic Penning-trap experiment reported in Ref.~\cite{Gabrielse:1990aa}), which is much lower than the $5\times10^{-13}$\,mbar considered in Fig.~\ref{fig:result}a,b). }} The trapping of the cylinder can be done  magnetically or electrically~\cite{Grosardt:2016aa, Armano:2016aa}, while the control and readout of the rotational motion can be achieved by an optical scattering technique \cite{Vovrosh:2017aa}.

Further, a stroboscopic detection mode can be chosen to suppress the heating by the detection light of the rotational motion of the cylinder~\cite{Goldwater:2016aa}. The rotational state can be prepared very reliably by feedback control~\cite{Kuhn:2017ac}. The feedback is then turned off to allow for heating. 

Alternatively, if a magnetic cylinder is levitated and trapped, SQUID sensors could be used to read the rotational state. The most notable advantage of this approach is that levitation can be achieved with static fields, implying negligible heat leak. Moreover, a temperature resolution better than 0.1\,K has already been demonstrated for state of the art high quality cantilevers with a SQUID-based magnetic detection \cite{Vinante:2011aa,Vinante:2016aa}. Therefore the main requirements for this proposal can be reached in a dedicated experiment based on existing technology. 

Clearly a big experimental challenge will be the control of seismic and acoustic noise and other environmental effects~\cite{Prat-Camps:2017ab}. In this respect, rotational degrees of freedom can be decoupled from vibrational noise much more effectively than vibrational ones. A well known and paradigmatic example{{ is given by the torsion pendulum}}, which is by far the most effective method to measure forces at Hz and sub-Hz frequencies.

Notice also that this non-interferometric test does not require the preparation of any non-classical state, which would need much advanced technology, yet to be demonstrated for such a macroscopic object. While the experimental scenario and the shape of the cylinder is the same as discussed already in Refs.~\cite{Collett:2003aa, Bera:2015aa}, here we find that macroscopic dimensions for the cylinder are useful for testing collapse models. This should make our proposed test far less demanding than experimenting with a nano- or micro-scale cylinder.\\

  {\section{Space-based Experiments}}

 An example of experiment in which rotational measurements can in principle improve the bounds on CSL with respect to translational tests is the space mission LISA Pathfinder, whose preliminary data for the vibrational noise were already exploited to set upper bounds on the CSL parameters \cite{Carlesso:2016ac,Helou:2017aa}. We can readily apply our model to LISA Pathfinder, with minimal variations to take into account the cubic instead of cylindrical geometry.
The core of the experiment consists in a pair of test masses in free-fall, surrounded by a satellite which follows the masses while minimizing the stray disturbances.  {As the satellite does not rigidly move with the masses, there is the necessity of setting a reference. One thus considers two masses in place of a single one, and focus on their relative motion.}
The geometry of each test mass is that of a cube of side $L=4.6$\,cm and mass $m=1.928$\,kg made of an AuPt alloy. The distance between the masses is $a=37.6$\,cm. 
Under the LISA Pathfinder conditions and provided that the noise is dominated by gas damping, in the limit of $r_C/L\to0$, we find that the torque over the force DNS ratio is 4 times bigger for the CSL noise than for the residual gas noise, showing that is advantageous to set bounds on CSL by looking at rotational noise. 
Though the data from the rotational noise measurements are not yet available, we can set an hypothetical bound on CSL parameters by
 converting the force DNS $S_\text{\tiny F}(\omega)=3.15 \times 10^{-30}$\,N$^2/$Hz \cite{Armano:2018aa} in torque DNS $\mathcal S_{\tau_x}(\omega)=0.04\times S_\text{\tiny F}(\omega)L^2=2.66\times 10^{-34}\,$N$^2$m$^2/$Hz.
Such a value is compared with the CSL contribution $\mathcal S_{\tau_x}(\omega)=\tfrac122\hbar^2\eta_\text{\tiny R}^\text{\tiny (cube)}$, where $\eta_\text{\tiny R}^\text{\tiny (cube)}$ is given in Eq.~\eqref{etaLISA} and the factors $\tfrac12$ and 2 account respectively for the differential measurement of the two masses of LISA Pathfinder and for the conversion from the two-side to one-side spectra. 
The corresponding upper bound and excluded region are shown in red in Fig.~\ref{fig:result}c). 
In blue, we report the upper bound from the vibrational analyses performed in \cite{Carlesso:2016ac} with the improved data from \cite{Armano:2018aa}.
The rotational upper bound would correspond to an enhancement of a factor 4 with respect to the improved vibrational bound, which is already almost one order of magnitude stronger than the one previously established \cite{Carlesso:2016ac}.
This shows that a rotation-based tests hold the potential to refine the probing of CSL mechanisms.

We underline that this bound is hypothetical, as the rotational noise is theoretically estimated. The measurement accuracy of the rotational motion is expected to be worse since the interferometric measurement of LISA Pathfinder is optimized for the vibrational degrees of freedom \cite{Weber:private}. To get a rotational bound, which is stronger than the vibrational one, one needs $\mathcal S_{\tau_x}(\omega)=1.07\times10^{-33}\,$N$^2$m$^2/$Hz. This should be within reach of the LISA Pathfinder technology. A more technical discussion that includes environmental noise is given in Appendix~\ref{app:LISA}.\\
A final note: the hypothetical rotational upper bound would completely rule out the possibility that the excess noise measured in the improved cantilever experiment \cite{Vinante:2017aa} is due to the CSL noise.\\

  {\section{Conclusions}}

The CSL parameter space has been the focus of a growing number of theoretical and experimental investigations aimed at reducing it significantly. To date, the region of the parameter space for this model that has not been excluded explicitly is still many orders of magnitude wide both in the values of the correlation length $r_C$ and the localization rate $\lambda$ (cf.~white region in Fig.~\ref{fig:bestgeom}a).
We have proposed a non-interferometric test capable of probing such a region. {The difference with the tests that have already been suggested and performed is twofold. First, our proposal is built on the use of rotational degrees of freedom rather than the usual vibrational ones. Second, the scheme focuses on objects of macroscopic dimensions instead of micro-scale ones. As discussed in detail in this work, both aspects offer considerable advantages that were at the basis of the reduction of the parameter space mentioned above. Although both features above have already been discussed and studied individually, an investigation combining such advantages together is unique of our proposal. We believe that the test that has been put forward here, which has been shown to adhere well to the current experimental state of the art, provides a new avenue of great potential for testing the CSL model. }\\

\noindent
\textit{Acknowledgments --} 
The authors acknowledge support from EU FET project TEQ (grant agreement 766900).
AB, MP, HU and AV acknowledge support from COST Action CA 15220 QTSpace.
AB acknowledges financial support from the University of Trieste (FRA 2016) and INFN.
MP acknowledges support from the DfE-SFI Investigator Programme (grant 15/IA/2864), and the Royal Society Newton Mobility (grant NI160057).
HU acknowledges financial support by The Leverhulme Trust (RPG-2016-046) and the Foundational Questions Institute (grant 2017-171363 (5561)). 
{{}{AV thanks WJ Weber for technical discussions on LISA Pathfinder.}}
This research was supported in part by the International Centre for Theoretical Sciences (ICTS) under the visiting program -- Fundamental Problems of Quantum Physics (Code: ICTS/Prog-fpqp/2016/11).

%


\appendix
\section{CSL Diffusion coefficients}
\label{AppB}

The CSL diffusion coefficients have been already computed in \cite{Bahrami:2014aa,Nimmrichter:2014aa,Schrinski:2017aa}. 
Given the mass density $\mu(\r)$, they read
\bqali\label{etavr}
\etat&=\frac{\lambda r_C^3}{\pi^{3/2}m_0^2}\int\Dd{3}\k\, e^{-r_C^2k^2}k_x^2|\tilde\mu(\k)|^2,\\
\etar&=\frac{\lambda r_C^3}{\pi^{3/2}m_0^2}\int\Dd{3}\k\, e^{-r_C^2k^2}|k_y\partial_{k_z}\tilde\mu(\k)-k_z\partial_{k_y}\tilde\mu(\k)|^2,
\eqali
where $\tilde\mu(\k)$ is the Fourier transform of $\mu(\r)$.
For a cylinder of length $L$ and radius $R$ we have \cite{Nimmrichter:2014aa,Schrinski:2017aa}
\begin{widetext}
{{}{
\bqali\label{etaexplicit}
\etat^\text{\tiny (cyl)}&=\frac{8m^2r_C^2\lambda}{ L^2m_0^2R^2}{\besseli_1\!\left(\tfrac{R^2}{2r_C^2}\right)}{e^{-\tfrac{R^2}{2r_C^2}}}\left(\tfrac{L\sqrt{\pi}}{2r_C}\erf(\tfrac{L}{2r_C})-1+e^{-\tfrac{L^2}{4r_C^2}}\right),\\
\etat^\text{\tiny (cyl, sym)}&=\frac{8m^2r_C^2\lambda}{L^2m_0^2R^2}\left(1-e^{-\tfrac{L^2}{4r_C^2}}\right)\left(1-e^{-\tfrac{R^2}{2r_C^2}}\left(\besseli_0\left(\tfrac{R^2}{2r_C^2}\right)+\besseli_1\left(\tfrac{R^2}{2r_C^2}\right)\right)\right),\\
\etar^\text{\tiny (cyl)}&=\frac{2\lambda r_C^4m^2}{L^2R^2m_0^2}\left\{\left[\left(1-e^{-\tfrac{L^2}{4r_C^2}}\right)\left(8+\frac{R^2}{r_C^2}\right)-\frac{2L\sqrt{\pi}}{r_C}\erf\left(\tfrac{L}{2r_C}\right)\right]\right.\\
&\left.-2\besseli_0\!\left(\tfrac{R^2}{2r_C^2}\right)e^{-\tfrac{R^2}{2r_C^2}}\left[\left(1-e^{-\tfrac{L^2}{4r_C^2}}\right)\left(4+\frac{R^2}{r_C^2}\right)-\frac{L\sqrt{\pi}}{r_C}\erf\left(\tfrac{L}{2r_C}\right)\right]\right.\\
&\left.-\tfrac13\besseli_1\!\left(\tfrac{R^2}{2r_C^2}\right)e^{-\tfrac{R^2}{2r_C^2}}\left[\left(3-e^{-\tfrac{L^2}{4r_C^2}}\right)\frac{L^2}{r_C^2}+2\left(1-e^{-\tfrac{L^2}{4r_C^2}}\right)\left(14+\frac{3R^2}{r_C^2}\right)-\frac{L\sqrt{\pi}}{2r_C}\left(24+\tfrac{L^2}{r_C^2}\right)	\erf\left(\tfrac{L}{2r_C}\right)\right]	\right\},
\eqali}}
where $\besseli_n$ denotes the $n$-th modified Bessel function. {{}{We also need the following coefficient
\bqali\label{etaLISA}
&\eta_\text{\tiny R}^\text{\tiny (cube)}=\frac{8\lambda}3 \left(	\frac{m}{m_0}\right)^2\left(\frac{r_C}{L}\right)^6 \left(1-e^{-\tfrac{L^2}{4r_C^2}}-\frac{\sqrt{\pi}L}{2r_C}\erf\left(\tfrac{L}{2r_C}\right)\right)\\
&\times\left\{	\left(1-e^{-\tfrac{L^2}{4r_C^2}}\right)\left[	2\left(3-e^{-\tfrac{L^2}{4r_C^2}}\right)\left(\frac{L}{r_C}\right)^2+32\left(1-e^{-\tfrac{L^2}{4r_C^2}}\right)-\frac{\sqrt{\pi}L}{r_C}\left[24+\left(\frac{L}{r_C}\right)^2\right]\erf(\tfrac{L}{2r_C})\right]+3\pi\left(\frac{L}{r_C}\right)^2 \erf^2(\tfrac{L}{2r_C})\right\},
\eqali
which refers to a cube of side $L$.}}
\end{widetext}

\section{Effective frequencies and damping constants}
\label{AppA}

The effective frequencies $\omegameff$ and $\omegapeff$ and damping constants $\gammameff$ and $\Dpeff$ introduced in Table \ref{tavola} take the following form

\bqali
\omegameff^2&=\omegam^2-\frac{2\hbar\chi^2|\alpha|^2\Delta(\kappa^2+\Delta^2-\omega^2)}{m\left(\kappa^2+(\Delta-\omega)^2\right)\left(\kappa^2+(\Delta+\omega)^2\right)},\\
\omegapeff^2&=\omegap^2-\frac{2\hbar\gp^2|\alpha|^2\Delta(\kappa^2+\Delta^2-\omega^2)}{I\left(\kappa^2+(\Delta-\omega)^2\right)\left(\kappa^2+(\Delta+\omega)^2\right)},
\eqali
\bqali
\gammameff&=\gammam+\frac{4\hbar\chi^2|\alpha|^2\kappa\Delta}{m\left(\kappa^2+(\Delta-\omega)^2\right)\left(\kappa^2+(\Delta+\omega)^2\right)},\\
\Dpeff&=\Dp+\frac{4\hbar\gp^2|\alpha|^2\kappa\Delta}{\left(\kappa^2+(\Delta-\omega)^2\right)\left(\kappa^2+(\Delta+\omega)^2\right)}.
\eqali

The damping constants $\gammam$ and $\Dp$ can be expressed in terms of the parameters of the system~\cite{Cavalleri:2010aa}
\bqali
\gammam&=\frac{P}{m}\sqrt{\frac{2\pi m_\text{\tiny gas}}{\kb T}} R^2\left[1+\frac{3L}{2R}\left(1+\tfrac{\pi}{6}\right)\right],\\
\Dp&=P\sqrt{\frac{\pi m_\text{\tiny gas}}{2\kb T}}R^4\left[1+\frac\pi4+\frac LR+\frac12\left(\frac LR\right)^2\right.\\&\left.+\frac14\left(\frac LR\right)^3\left(1+\frac \pi6\right)\right],
\eqali
where $P$ is the pressure of the surrounding gas of particles of mass $m_\text{\tiny gas}$. For the vibrational motion along the symmetry axis the damping rate $\gammam$ must be substituted by the following expression \cite{Cavalleri:2010aa}
\bqali
\gammam^\text{\tiny sym}&=\frac{P}{m}\sqrt{\frac{8\pi m_\text{\tiny gas}}{\kb T}} R^2\left(1+\frac{\pi}{4}+\frac{L}{2R}\right).
\eqali
This gives the green lines in Fig.~\ref{fig:bestgeom}.

{{}{
\section{Analysis of LISA Pathfinder noises}
\label{app:LISA}

Whether one can set stronger bounds on collapse models by looking at the translational or the rotational noise of a given mechanical system depends crucially on the specific experimental implementation. In particular, one has to compare how the CSL noise and the dominant (physical) residual noise scale when passing from translational to rotational noise. If the scaling is different, the bounds that can be inferred from the same experimental setup under the same conditions are different. Here, we show that, for the specific experiment of LISA Pathfinder, under the assumption that residual gas is the dominant source of noise, rotational noise is in principle the best choice.
We limit our analysis to the short CSL length limit $r_C \ll L$, which is the relevant one in the case of LISA.

We introduce a dimensionless factor $\alpha$, defined as:
\begin{equation}
\alpha _i L^2  = \frac{{\mathcal S_{\tau ,i} }}{{\mathcal S_{F,i} }},
\end{equation}
where $\mathcal S_{\tau ,i}$ and $\mathcal S_{F,i} $ are the torque and force DNS respectively, and the pedices $i$ may refer to the three specific cases `CSL', `gas-$\infty$' and `gas' which we will now discuss. For the residual gas noise we consider both the case of gas within an infinite volume (\emph{gas-$\infty$}) and the real case of a gas constrained in a small gap $d\ll L$ (\emph{gas}), which is the relevant one for LISA Pathfinder. Essentially, $\alpha$ is the effective ratio of rotational (torque) noise to vibrational (force) noise for a given source.

For CSL in the case of a cubic mass with $r_C/L\rightarrow 0$, comparison of Eq.~\eqref{etaLISA} with the formula for the vibrational diffusion constant \cite{Vinante:2016aa,Nimmrichter:2014aa}:
\bq
\eta_\text{\tiny V}^\text{\tiny (cube)}=\frac{32 \lambda r_C^4 m^2}{m_0^2 L^6}(\tfrac{\sqrt{\pi}L}{2r_C}\erf(\tfrac{L}{2r_C})-1+e^{-\tfrac{L^2}{4r_C^2}})^2(1-e^{-\tfrac{L^2}{4r_C^2}}),
\eq
in the same limit, provides $\alpha_{\mathrm CSL}=1/6 \simeq 0.166$. This factor is the same as for a gas of uncorrelated particles scattering elastically off the test mass, which is the typical picture considered in elementary textbooks of statistical mechanics. In fact, under elastic scattering the force exerted by the gas is normal to the surface and the total force noise is proportional to the exposed area. One can write $\D S_F=a\cdot \D A$, and $\D S_\tau= a b^2\cdot \D A$, where $a$ is the noise strength and $b$ is arm of the force, i.e.~the distance between the normal to the surface at a given point and the rotational axis. Elementary integration of the force and torque on each cube face provides precisely the factor $\alpha = 1/6$ regardless of the direction of the force and the torque. In this sense, CSL in the $r_C\ll L$ limit behaves essentially as a gas of uncorrelated particles hitting the surface normally, which can be interpreted as a collection of uncorrelated collapse events localized on the cube surface. 

For a gas in a real experiment, the elastic scattering assumption is known to be wrong. A vast experimental evidence suggests that the data are instead consistent with the inelastic diffuse scattering model \cite{Cavalleri:2010aa}. In this picture, a particle hitting the surface with a given angle $\theta_i$ is reflected with a different angle $\theta_r$, with joint probability proportional to $\mathrm{cos} \theta_i \times \mathrm{cos }\theta_r$ and with uncorrelated incidence and emission velocities consistent with a Maxwell-Boltzmann distribution. In general a shear force component will appear in addition to the normal component. Detailed calculations have been carried out analytically for a gas of particles within an infinite volume \cite{Cavalleri:2010aa}. For a cube it is found that the ratio of rotational to vibrational noise is $\alpha_{\mathrm{gas}-\infty } \simeq 0.226 > \alpha_{\mathrm CSL}$. Therefore, for infinite volume it is not advantageous to set bounds on CSL by looking at rotational noise.

However, the situation of LISA Pathfinder is slightly more complex. The cubic test mass (TM) is enclosed in an external caging, the gravitational reference sensor (GRS), with a relatively narrow gap between TM and GRS, in the range $d/L =[ 0.063, 0.087] \ll 1$ depending on the axis. The gap is narrow in order to enable continuous monitoring (with subsequent control of the GRS) of the relative position between GRS and TM by means of capacitive electrodes.

Under the gap constraint, each individual gas particle inside the gap will undergo a random walk with a large number of multiple collisions, introducing a degree of correlation between consecutive events. Extensive investigation of this effect has been carried out, based on numerical simulations and experiments with torsion pendulums \cite{Cavalleri:2009aa, Dolesi:2011aa}. The multiple scattering is found to introduce a correlation time $\tau_c$, related to the mean time required by a particle to random-walk along the gap from one side to the other side of a cubic face. At frequencies $\omega \tau_c \ll  1$, the relevant case for LISA Pathfinder, there is a significant increase of both the vibrational force and rotational noise, and related damping factors. In the limit $d\ll L$ the increase goes asymptotically as $(L/d)^2$. 

It turns out that the increase of vibrational noise is significantly larger than the increase of rotational noise. This can be intuitively understood as following. When a gas particle, during its random walk, crosses the center of a cubic face, the sign of the torque changes whereas the sign of the force does not. As a consequence, the correlation time for the rotational case is shorter by a factor around 4, being related with the time required to random-walk along half of the cubic face instead of the whole face.

A quantitative estimation can be done by inspection of Fig. 5 in Ref.~\cite{Cavalleri:2009aa}. Under the LISA Pathfinder condition, the force noise will increase with respect to the infinite volume case by a factor roughly 6 times larger than the torque noise. This leads to $\alpha_{\mathrm{gas}} \simeq  0.04 < \alpha_{\mathrm{CSL}}$.
Thus, provided that the noise in LISA Pathfinder is dominated by gas damping, rotational noise allows to set an upper bound on CSL which is about $4$ times better than using vibrational noise. This is shown in Fig.~\ref{fig:result}c) of the main text.

It is worth mentioning a final aspect. Rotational measurements in LISA Pathfinder are slightly more complicated than vibrational ones. In fact, the residual rotational noise is obtained from a differential measurement scheme which provides a cancellation of the actuation noise \cite{Weber:private}. As the latter is two orders of magnitude larger than the former, a very accurate calibration of the actuation noise is needed. On the other hand, an imperfect cancellation of the actuation noise will unavoidably degrade the rotational noise spectrum. This might eventually spoil the advantage of using rotational noise to set an ultimate bound on CSL with LISA Pathfinder. This issue could be solved in a dedicated similar experiment in which the readout is designed to directly measure rotations instead of vibrations.
}}

\end{document}